\documentclass[aps,prd,twocolumn,showpacs,amsmath,amssymb]{revtex4}

\usepackage{slashed}
\usepackage{graphicx} 
\usepackage{dcolumn} 
\usepackage{bm} 
\usepackage{epsfig}
\usepackage{amsmath}
\usepackage{amssymb}
\usepackage{hyperref}
\usepackage{color}
\usepackage{ulem}

\begin{document}

\newcommand{\bsy}[1]{\mbox{${\boldsymbol #1}$}}
\newcommand{\bvecsy}[1]{\mbox{$\vec{\boldsymbol #1}$}}
\newcommand{\bvec}[1]{\mbox{$\vec{\mathbf #1}$}}
\newcommand{\btensorsy}[1]{\mbox{$\tensor{\boldsymbol #1}$}}
\newcommand{\btensor}[1]{\mbox{$\tensor{\mathbf #1}$}}
\newcommand{\tensorId}{\mbox{$\tensor{\mathbb{\mathbf I}}$}}
\newcommand{\be}{\begin{equation}}
\newcommand{\ee}{\end{equation}}
\newcommand{\bea}{\begin{eqnarray}}
\newcommand{\eea}{\end{eqnarray}}

\title{The relativistic electron gas: a candidate for nature's left-handed material}

\author{C. A. A. de Carvalho$^{1,2}$}

\affiliation{$^{1}$Instituto de F\'{\i}sica, Universidade Federal do Rio de Janeiro - UFRJ, Caixa Postal  68528, Rio de Janeiro, RJ, 21945-972, Brasil\\
$^{2}$Instituto Nacional de Metrologia, Qualidade e Tecnologia - INMETRO, Av. Nossa Senhora das Gra\c cas, 50, Xer\' em, Duque de Caxias, 25250-020, Brasil}

\date{\today}

\begin{abstract}
The electric permittivities and magnetic permeabilities for a relativistic electron gas are calculated from quantum electrodynamics at  finite temperature and density as functions of temperature, chemical potential, frequency, and wavevector. The polarization and the magnetization depend linearly on both electric and magnetic fields, and are the sum of a zero-temperature and zero-density vacuum part with a temperature- and chemical potential-dependent medium part. Analytic calculations lead to generalized expressions that depend on three scalar functions. In the nonrelativistic limit, results reproduce the Lindhard formula. In the relativistic case, and in the long wavelength limit, we obtain: i)  for $\omega=0$, generalized susceptibilities that reduce to known nonrelativistic limits; ii) for $\omega \neq 0$, Drude-type responses at zero and at high temperatures. The latter implies that one may have both the eletric permittivity $\epsilon$ and the magnetic permeability $\mu$  simultaneously negative, a behavior characteristic of metamaterials. This unambiguously indicates that the relativistic electron gas is one of nature's candidates for the realization of a negative index of refraction system. Moreover, Maxwell's equations in the medium yield the dispersion relation and the index of refraction of the electron gas. Present results should be relevant for plasma physics, astrophysical observations, synchrotrons, and other environments with fast moving electrons.
\end{abstract}


\maketitle


Metamaterials  evolved from a mathematical curiosity \cite{Veselago} to real applications thanks to the nanoengineering that made them accessible in the laboratory \cite{Smith}. As a consequence, they rekindled interest in the electromagnetic responses of material media and led us to search for a  reliable way to calculate such responses for a relativistic electron gas. Thus, we resorted to quantum field theory \cite{Early,Recent,KG,Texts2} to investigate a system of relativistic electrons at finite temperature and density. Besides providing a test of the treatment, that system is a reasonable approximation to physical situations encountered in plasma physics, astrophysics, synchrotrons, and other environments with fast moving electrons.

One should note that no examples of naturally occurring materials with $\epsilon$ electric permittivity and $\mu$ magnetic permeability simultaneously negative were ever found \cite{Pendry}. As there seems to be no reason why that could not happen in nature, we have decided to look for such effects in the relativistic electron gas at finite density and finite temperature for which analytic studies of electromagnetic responses may be straightforwardly performed by using a quantum field theoretical approach. Indeed, as we shall demonstrate, for long-wavelength radiation, a finite density of relativistic electrons exhibits Drude-type responses for both $\epsilon$ and $\mu^{-1}$, at $T=0$, as well as at high temperatures, implying that they can be simultaneously negative for frequencies that are low when compared to the electric plasmon frequency.

Let us first consider the partition function  $Z=\hbox{Tr} \, e^{-\beta (\hat{H}-\xi\Delta \hat{N}) }$ of quantum electrodynamics (QED) at finite temperature and density, which describes an electron gas with fixed $\Delta N= N_e - N_p$ ($N_e$ is the number of electrons; $N_p$ is the number of positrons) at temperature $T=\beta^{-1}$ (Boltzmann constant $k_B =1$) and chemical potential $ \xi$, coupled to the electromagnetic field $A_\nu$. $Z$ may be expressed as a functional integral over gauge and fermion fields  \cite{KG}
\be
\label{Z}
Z=\oint  [d\Omega] \, \delta({\cal G}) e^ {-S_{A}[A]} Z_{e}[A]\,\,,
\ee
where
\be
Z_e [A] = \oint [i d\psi^{\dagger}][d\psi] e^{- S_{e} [\psi^{\dagger}, \psi, A]} \,\,,
\ee
$[d\Omega] \equiv [dA_\nu] \det (\delta {\cal G}/\delta \Lambda)$, and the determinant is the Jacobian of the gauge transformation $A_\nu\rightarrow A_\nu - \partial_\nu \Lambda$. Notice that the delta function imposes the gauge condition ${\cal G}[A]=0$, typically ${\cal G}[A]=\partial_\nu A_\nu$, the actions $S_X=  \int_0^{\beta} dx_4  \int d^3 x   \,{\cal L}_X$ $(X = A,\,e)$ involve ${\cal{L}}_{A}= -\frac{1}{4} F_{\mu\nu} F_{\mu\nu}$ and ${\cal L}_e=  \bar{\psi}  \Gamma_{A} \psi $, $\Gamma_{A}=G_{A} ^ {-1} = i \slashed{D} - m - i \xi\gamma_4 $ is the inverse of the electron propagator in the presence of the gauge field, $\slashed{D} \equiv \gamma . (\partial  - ie A)$, $e$ and $m$ are the electron charge and mass, and $\hbar=1, c=1$, $\bar{\psi}= \psi^{\dagger} \gamma_4$.  The integral $\oint$ runs over gauge fields obeying $A_\nu(0,\vec{x})=A_\nu(\beta, \vec{x})$, and electron fields obeying $\psi (0,\vec{x})=- \psi(\beta, \vec{x})$ \cite{MtoE}. 

In the lowest order of a semiclassical approximation (see Appendix \ref{App1}), we take $A_\mu$ to be a classical field, and integrate over the electron field, to obtain
\be
\label{3}
Z_e[A]=\det [-\beta \gamma_4 \Gamma_{A}] = \exp \hbox{Tr} \ln [-\beta \gamma_4 \Gamma_{A}]. \\
\ee
This leads to a modified action for the $A_\mu$ field, $S_{sc}[A] = S_{A}[A] -Tr \ln [ -\beta \gamma_4 \Gamma_{A}]$, which takes into account the response of the electrons. The extremal condition $\delta S_{sc}/ \delta A_{\nu} = 0$ gives the equation of motion
\be
\partial_\mu F_{\mu\nu} = - Tr[e \gamma_\nu G_A] = J_\nu.
\ee
Splitting $J$ into free $J^{(0)}$ (for $ A=0$) and induced $J^{(I)}$ currents, we may rewrite the equation of motion as
\bea
&& \partial_\mu ( F_{\mu\nu} +  P_{\mu\nu})= J_\nu^{(0)}, \\
&& -\partial_\mu P_{\mu\nu}= J_\nu^{(I)}= Tr[e \gamma_\nu G_A] - Tr[e \gamma_\nu G_0],
\eea
with $G_0$ the free electron propagator. $P_{\mu\nu}$ defines the polarization  $\vec P$ ($P_{4j}= iP^j$)  and magnetization $\vec M$ ($P_{ij}= - \epsilon_{ijk} M^k$) vectors. Expanding the current in the field $A_\nu$ yields an infinite series of one-loop graphs, which is equivalent to the the RPA approximation \cite{PN} of Condensed Matter Physics. If we only retain the linear term, which amounts to the linear response approximation, we obtain the momentum space equation
\be
i q_\mu \tilde{P} _{\mu\nu}(q)= \tilde{\Pi}_{\nu\sigma}(q) \tilde{A}_\sigma(q),
\label{PPiA}
\ee
where
\be
\tilde{\Pi}_{\nu\sigma} = - \frac {e^2}{\beta} \sum_{n=-\infty} ^ {+\infty} \int \frac {d^3 p}{(2\pi)^3} \hbox{Sp} [\gamma_\nu G_0(p) \gamma_\sigma G_0 (p-q)].
\label{pol}
\ee
The sum is over Matsubara frequencies $p_4= (2n+1) \pi T$, with Sp denoting trace over Dirac matrices.  The solution to Eq. \eqref{PPiA},
\be
\tilde{P}_{\mu\nu} = \frac{\tilde{\Pi}_{\mu\sigma}}{q^2} F_ {\nu\sigma}- \frac{\tilde{\Pi}_{\nu\sigma}}{q^2} F_{\mu\sigma},
\label{Pmunu}
\ee
relates polarization and magnetization to the fields $\vec{E}$  ($F_{4j}= iE^j$)  and $\vec B$ ($F_{ij}= \epsilon_{ijk} B^k$), thus yielding electric and magnetic susceptibilities and,  ultimately, electric permittivities and magnetic permeabilities. One may write $\tilde{\Pi}_{\nu\sigma}=\tilde{\Pi}_{\nu\sigma}^{(v)} + \tilde{\Pi}_{\nu\sigma}^{(m)}$ to separate vacuum ($T= \xi = 0$) and medium contributions. The vacuum gives
\be
 -\frac{\tilde{\Pi}_{\nu\sigma} ^{(v)}}{ q^2 }= (\delta_{\nu\sigma} - \frac{q_\nu q_\sigma}{q^2}) {\cal{C}}(q^2).
\label{vac}
\ee
The medium, however, introduces a preferred reference frame (that of its center of mass). The symmetry is then reduced to three-dimensional rotation and gauge invariance, leading to

\bea
\label{med1}
&& -\frac{\tilde{\Pi}_{ij} ^{(m)}}{ q^2 }= (\delta_{ij} - \frac{q_iq_j}{|\vec{q}|^2}) {\cal{A}} + \delta_{ij} \frac{q_4^2}{|\vec{q}|^2} {\cal{B}},  \\
&& -\frac{\tilde{\Pi}_{44} ^{(m)}}{q^2 } = {\cal B},  \,\,\,\,\,\,\,\,\,  -\frac{\tilde{\Pi}_{4i} ^{(m)}}{q^2 } = - \frac{q_4 q_i}{|\vec{q}|^2} {\cal B},
\label{med2}
\eea
where ${\cal{A}} (q_4, |\vec{q}|)$, ${\cal B} (q_4, |\vec{q}|)$, and ${\cal{C}}(q^2)$ are determined from the Feynman graph in Eq. \eqref{pol}, which corresponds to the QED polarization tensor at finite temperature and density, computed long ago \cite{AP}. ${\cal{A}}$ and ${\cal{B}}$ are calculated from the trace $\tilde{\Pi}_{\mu\mu}$ and $\tilde{\Pi}_{44}$, once we subtract the vacuum part, i.e,
\begin{equation}
{\cal{A}}= \frac{-e^{2}}{2 \pi^{3} q^{2}} \mathrm{Re} \int \frac{d^3 p}{\omega_p} n_{F} (p) \frac{p.(p+q)}{q^2-2p.q} + (1- \frac{3q^2}{2|\vec{q}|^2}){\cal{B}},
\label{calA}
\end{equation}
and
\begin{equation}
{\cal{B}}=\frac{-e^{2}}{2 \pi^{3} q^{2}} \mathrm{Re} \int \frac{d^3 p}{\omega_p} n_{F} (p) \frac{p.q - 2p_4(q_4 - p_4)}{q^2-2p.q} ,
\label{calB}
\end{equation}
where $p_4=i\omega_p=i \sqrt{|\vec{p}|^2+ m^2}$ and $n_{F}(p)= (e^{\beta(\omega_p - \xi)} +1)^{-1} +  (e^{\beta(\omega_p + \xi)} +1)^{-1} $.
Expressions \eqref{calA} and \eqref{calB}  may be integrated  over angles \cite{AP} (Appendix \ref{App2}). ${\cal{C}}$ is obtained from the vacuum polarization contribution \cite{IZ}.

The Euclidean space $\tilde{\Pi}_{\mu\nu}$ is a function of the Euclidean $q_4$. As  ${\Pi}_{\mu\nu}(x,y)$ may be expressed as a current-current correlation $\langle j_\mu(x)\, j_\nu(y)\rangle$, $j_\mu=\psi^{\dagger}\gamma_4\gamma_\mu\psi$, it may be written in terms of a spectral density  $\tilde{\rho}_{\mu\nu}$
\be
\label{SD}
\tilde{\Pi}_{\mu\nu} (\omega_n, \vec{q})=\int_{-\infty}^{+\infty} \frac{dv} {v+ i \omega_n}{\tilde{\rho}_{\mu\nu} (v, \vec{q})},
\ee
where $\tilde{\rho}_{\mu\nu}$ is expressible in terms of expectation values of the eigenstates of the Hamiltonian. In order to obtain a Minkowski space expression, we need the current-current correlation for $j^\mu=\psi^{\dagger}\gamma^0\gamma^\mu\psi$, with the corresponding spectral density also given by expectation values \cite{KG,Texts2}. Then, using our conventions for the relation between Euclidean and Minkowski indices, we derive
\be
\label{EuMi1}
\tilde{\Pi}_{44}^{\ast} = i \tilde{\Pi}^{00}; \,\,\,\,\tilde{\Pi}_{4k}^{\ast} = \tilde{\Pi}^{0k}; \,\,\,\,\tilde{\Pi}_{kl}^{\ast} = -i \tilde{\Pi}^{kl},
\ee
where the asterisk means $q_ 4= \omega_n \rightarrow i\omega - 0^+$. Since Euclidean $q^2=q_4^2+|\vec{q}|^2$ goes to Minkowski $-q^2=-q_0^2+|\vec{q}|^2$, our prescription takes Euclidean quantities into the physical Minkowski ones through 
\be
\label{EuMi2}
\frac{\tilde{\Pi}_{44}}{q^2} \rightarrow \frac{-i \tilde{\Pi}^{00}}{q^2}; \frac{\tilde{\Pi}_{4k}}{q^2} \rightarrow \frac{- \tilde{\Pi}^{0k}}{q^2}; \frac{\tilde{\Pi}_{jk}}{q^2} \rightarrow \frac{i \tilde{\Pi}^{jk}}{q^2},
\ee
leading [cf.  Eq. \eqref{Pmunu}] to the Minkowski expressions
\begin{eqnarray}
\label{Pj}
&& \tilde{P}^{j} =  \frac{i{\tilde \Pi}^{00}}{q^2} \tilde{E}^{j} - \frac{i\tilde \Pi^{jk}}{q^2} \tilde{E}^{k} + i \epsilon^{jkl}\frac{\tilde \Pi^{0k}}{q^2} \tilde{B}^{l}, \\
&& \tilde{M}^{j} = \frac{i{\tilde \Pi}^{kk}}{q^2} \tilde{B}^{j} - \frac{i\tilde \Pi^{jk}}{q^2} \tilde{B}^{k} - i \epsilon^{jkl}\frac{\tilde \Pi^{0k}}{q^2} \tilde{E}^{l}.
\label{Mj}
\end{eqnarray}

We now introduce $H_{\mu\nu} =F_{\mu\nu}+ P_{\mu\nu}$, which defines $H_{4j} = iD^j$ and $H_{ij}= \epsilon_{ijk} H^k$, with $\vec{D}= \vec{E} + \vec{P}$ and $\vec{H} = \vec{B} - \vec{M}$. The constitutive equations are derived from Eqs. \eqref{vac}, \eqref{med1}, and \eqref{med2}, and Eqs. \eqref{Pj} and \eqref{Mj}, i.e,
\begin{eqnarray}
\label{Dconst}
&& \tilde D^j=\epsilon^{jk} \tilde E^k + \tau^{jk} \tilde B^k, \\
&& \tilde H^j= (\mu^{-1})^{jk} \tilde B^k + \sigma^{jk} \tilde E^k \, \, ,
\label{Hconst}
\end{eqnarray}
where, by using $\hat{q}^i\equiv q^i/|\vec{q}|$, one obtains
\bea
\label{resp1}
&& \epsilon^{jk}= \epsilon \delta^{jk} + \epsilon' \hat{q}^j \hat{q}^k, \\
&& (\mu^{-1})^{jk}= \mu^{-1} \delta^{jk} + \mu'^{-1} \hat{q}^j \hat{q}^k, \\
&& \tau^{jk}= \tau \epsilon^{jkl} \hat{q}^l, \,\, \sigma^{jk}= \sigma \epsilon^{jkl} \hat{q}^l \, \, .
\label{resp4}
\eea
 One should stress that there are contributions  to $(\vec{D}, \vec{H})$ along the directions of the fields $(\vec{E}, \vec{B})$, of the wavevector $\vec{q}$, and of $(\vec{q}\wedge\vec{B}, \vec{q}\wedge\vec{E})$; also note that bianisotropic crystals satisfy similar relations \cite{Bi}.

The permittivities and permeabilities
\bea
\label{responses1}
&& \epsilon=1+(2 - \frac{\omega^2}{q^2} ){\cal C}^\ast+{\cal A}^\ast+( 1- \frac{\omega^2}{|\vec{q}|^2}) {\cal B}^\ast, \\
\label{responses2}
&& \mu^{-1}=1+(2+ \frac{|\vec{q}|^2}{q^2}){\cal C} ^\ast+ {\cal A}^\ast - 2\frac{\omega^2}{|\vec{q}|^2} {\cal B}^\ast, \\
&& \epsilon'= - \mu'^{-1} = \frac{|\vec{q}|^2}{q^2}{\cal C}^\ast - {\cal A}^\ast, \\
\label{responses4}
&& \tau= \sigma= \frac{\omega}{|\vec{q}|}( \frac{|\vec{q}|^2}{q^2} {\cal C}^\ast - {\cal B}^\ast),
\eea
are determined by three scalar functions ${\cal A}^\ast$, ${\cal B}^\ast$, and ${\cal C}^\ast$, where again the asterisk means $q_ 4 \rightarrow i\omega - 0^+$ .  ${\cal C}^\ast$ may be obtained from the standard calculation at $T=\xi=0$ \cite{IZ}
\be
{\cal C}^\ast = \frac{-e^2}{12 \pi^{2}}\{ \frac{1}{3}+ 2(1+\frac{2m^2}{q^2}) [h \, \mathrm{arccot} (h) -1] \}
\ee
where $h=\sqrt{(4m^2/q^2) -1}$ and the renormalization condition is $e^2/(4\pi\hbar c)=1/137$, with $e^2=e^2(\omega=0,\vec{q}=\vec{0})$. The vacuum contributions to permittivities and permeabilities are obtained by setting ${\cal A}^\ast= {\cal B}^\ast=0$. On the other hand, medium susceptibilities may be defined as $\chi_e=\epsilon - \epsilon_{v}$; $\chi'_e=\epsilon' - \epsilon'_{v}$; $\chi_{em}=\tau - \tau_{v}$; $\chi_m=-(\mu^{-1} - \mu^{-1}_{v})$; $\chi'_m = -(\mu'^{-1} - \mu'^{-1}_{v})$; $\chi_{me}=-(\sigma - \sigma_{v})$. In the sequel, we shall examine  the long-wavelength limit $|\vec{q}| \rightarrow 0$ of nonrelativistic and relativistic expressions, for $\omega=0$ (the stationary case) and $\omega\neq 0$, for various physical quantities of interest.

Nonrelativistic expressions \cite{Texts2,Texts1} will follow whenever $|\xi - m| << m$, $\beta m << 1$,  $n_{F} \rightarrow n'_F= (e^{\beta(\varepsilon_p - \xi')} + 1)^{-1}$, with $\varepsilon_{\vec{p}}= |\vec{p}|^2/ 2m$ and $\xi'= \xi - m$. In Euclidean space, $q_4/m \sim T/m$, which leads to ${\cal A} \rightarrow 0$ and $|\vec{q}|^2/m^2 \sim T/m$. Then, ${\cal A}^\ast$ vanishes and $ \chi_{e} \rightarrow {\cal B}^\ast$ reduces to
\be
\chi_e \rightarrow {\cal B}^\ast = \frac{-2e^2}{ |\vec{q}|^2} \mathrm{Re} \int \frac{d^3 p}{4\pi^3} \frac{n'_F(\vec{p})}{\varepsilon_{\vec{p}} - \varepsilon_{\vec{p}-\vec{q}} - \omega-i0^+} \, \, .
\ee
The above expression may easily be converted into the Lindhard expression for the electric susceptibility,
\be
\chi_{e}\rightarrow {\cal B}^\ast =\frac{-e^2}{ |\vec{q}|^2} \mathrm{Re} \int \frac{d^3 p}{4\pi^3} \frac{ n'_{F} (\vec{p}+\vec{q})- n'_F(\vec{p})}{\varepsilon_{\vec{p}+\vec{q}} - \varepsilon_{\vec{p}} -\omega-i0^+} .
\ee
If one sets $\omega=0$ and goes to the long-wavelength limit
\be
\label{BNR}
{\cal B}^* = \frac{e^2 m}{\pi^2 |\vec{q}|^2}\int_0^\infty dp\, n'_F(p),
\ee
which is the Thomas-Fermi expression $\chi_e= m_{TF}^2/|\vec{q}|^2$, with
\be
\label{NRTF}
m_{TF}^2= \frac{e^2m}{\pi^2} \int_0^\infty dp \, n'_F (p).
\ee
If, instead, $\omega \neq 0$, in the long-wavelength limit, one obtains the Drude expression $\epsilon=1- \frac {\omega_{e}^2} {\omega^2}$, where the electric plasmon frequency is
\be
\omega_{e}^2 = \frac{-e^2}{\pi^2 m}\int_0^\infty dp\, p^2 \, n'_F(p).
\ee
For $T=0$, with $p_F= \sqrt {2m \xi'}$, and the electron density $n= p_F^3/3\pi^2$ ($\hbar=1$),
\be
\label{epl}
\omega_e^2 = \frac{e^2 m^2}{3\pi^2} \left(\frac{p_{F}}{m}\right)^3 = \frac{ne^2}{m}.
\ee
Since, in the nonrelativistic limit, ${\cal A} \rightarrow 0$ and $q_4/m \sim |\vec{q}|^2/m^2 \sim T/m$,  all other medium suceptibilities vanish in lowest order.

The relativistic case sets in as $|\xi - m|\sim m$, or $T \sim m$, or both. Then, generalized expressions must be used to investigate the stationary case, $\omega=0$, as well as the case $\omega \neq 0$, in the long wavelength limit $|\vec{q}| \rightarrow 0$. In the stationary case $\omega=0$, one has $\chi_e={\cal A}^\ast + {\cal B}^\ast$, $\chi'_e=\chi_{m}= \chi'_m = -{\cal A}^\ast$. In the long-wavelength limit,
\bea
\label{zeromegaA}
&& {\cal A}^\ast = - \frac{e^2}{6\pi^2} \int_0^\infty \frac{dp \, n_F (\vec{p})}{\sqrt{|\vec{p}|^2+m^2}} \, \, , \\
&& {\cal B}^\ast = \frac{e^2}{\pi^2 |\vec{q}|^2} \int_0^\infty \frac{dp \, n_F (\vec{p})}{\sqrt{|\vec{p}|^2+m^2}}  ( m^2+\frac{3}{2} |\vec{p}|^2).
\label{zeromegaB}
\eea
For $T=0$, one obtains a closed relativistic expression ($\xi/m\equiv \zeta$),
\be
\chi'_e=\chi_{m}= \chi'_m =  \frac{e^2}{6\pi^2} \mathrm{arccosh}(\zeta).
\ee
In the nonrelativistic limit,
\be
\chi'_e=\chi_{m}= \chi'_m = \frac{e^2}{6\pi^2} \frac{\sqrt {2m \xi'}}{m}= \frac{e^2}{6\pi^2} \frac{p_{F}}{m},
\ee
where ${p_F=\sqrt{2m\xi'}}$ is the Fermi momentum. That is just
\be
\chi'_e=\chi_{m}= \chi'_m= \frac{e^2}{4\pi^2}\frac{p_F}{m} - \frac{e^2}{12\pi^2}\frac{p_F}{m},
\ee
the sum of $\chi_{Pauli}= (e^2/4\pi^2 \hbar c) (p_F/mc)$ and $\chi_{Landau}= -(e^2/12\pi^2 \hbar c) (p_F/mc)$, where we have restored the units $\hbar$ and $c$. For $\chi_e$, the long-wavelength limit yields the relativistic generalization of the Thomas-Fermi expression $\chi_e= m_{TF}^2/|\vec{q}|^2$, with
\be
m_{TF}^2= \frac{e^2}{\pi^2} \int_0^\infty \frac{dp \, n_F (\vec{p})}{\sqrt{|\vec{p}|^2+m^2}}  ( m^2+\frac{3}{2} |\vec{p}|^2).
\ee
For $T=0$, one finds
\be
\frac{m_{TF}^2}{m^2}= \frac{e^2}{4\pi^2} [ \mathrm{arccosh}(\zeta) + 3 \zeta \sqrt{\zeta^2 - 1}],
\ee
with nonrelativistic limit $m_{TF}^2/m^2= (e^2/\pi^2) \sqrt{2\xi'/m}$, or $m_{TF}^2/m^2= (e^2/\pi^2\hbar c)(p_F/mc)$, which is the same that we obtain from Eq. \eqref{NRTF} at $T=0$. 

In order to access the long-wavelength relativistic limit for $\omega \neq 0$, one takes $|\vec{q}| \rightarrow 0$ and expands the expressions for ${\cal A}^\ast$ and ${\cal B}^\ast$ after the angular integration (Appendix \ref{App2}). Note that the last terms in Eqs. \eqref{responses1}, \eqref{responses2} now dominate, whereas they do not contribute in leading order in the nonrelativistic regime. In fact, for nonrelativistic systems in thermal equilibrium with the radiation, we must have $\omega \sim  (|\vec{q}|^2/2m) << m$, so that  $(\omega^2/|\vec{q}|^2)<< 1$. Taking the long wavelength limit after the nonrelativistic one yields a Drude expression for $\epsilon$, but not for $\mu^{-1}$.

For $T=0$, the $\epsilon$ electric response is given by a Drude-type expression
\bea
&& \epsilon= 1 - \frac{\omega_{e}^2}{\omega^2}+ \frac{e^2}{3\pi^2} \, g_e(\zeta)+ O(\frac{\omega^2}{4m^2}), \\
&& \frac{\omega_{e}^2}{4m^2}=  \frac{e^2}{12 \pi^2} \frac{{(\zeta^2- 1)}^{3/2}}{\zeta}, 
\eea
and so is the $\mu^{-1}$ magnetic response,
\bea
&& \mu^{-1}= 1 - \frac{\omega_{m}^2}{\omega^2}- \frac{5e^2}{6\pi^2}\, g_m(\zeta)+ O(\frac{\omega^2}{4m^2}), \\
&& \frac{\omega_{m}^2}{4m^2}=  \frac{2 e^2}{12 \pi^2} \frac{{(\zeta^2- 1)}^{3/2}}{\zeta}, 
\eea
where $g_e$ and $g_m$ are given in the Appendix. We note that there is a relation between $\omega_{m}$ and $\omega_e$, i.e, $\omega_{m}=\sqrt{2} \,  \omega_{e}$, and that the vacuum contribution is $O(\omega^2/4m^2)$. The electron plasmon frequency $\Omega_e$ is defined as the zero of $\epsilon$, $\Omega_e^2= \omega_e^2 [1+ \frac{e^2}{3\pi^2} g_e(\zeta)]^{-1} \simeq \omega_e^2$. The $\zeta \rightarrow 1$ limit of the electric plasmon frequency  coincides with the one given in Eq. \eqref{epl}.

The Drude-type expressions imply that the electric and magnetic responses may be {\it simultaneously negative} for small $\omega$. We should emphasize that the Drude-type behavior comes solely from the medium contribution; the vacuum part does not exhibit any singular behavior.

One might wonder about the effect of corrections on the Drude result. Those coming from the interaction of electrons with the classical fields will be nonlinear in the fields, typically of order $\alpha(\alpha E^2/m^4)$ or $\alpha(\alpha B^2/m^4)$. Those coming from e-e interactions will have a linear response term of order $\alpha^2$ and nonlinear contributions also of order $\alpha(\alpha E^2/m^4)$, $\alpha(\alpha B^2/m^4)$. For fields that are not strong enough to invalidate the linear response approximation, those corrections will, presumably, be small compared to the leading contribution, and therefore unable to cancel the $1/\omega^2$ terms. 

Finally, we investigate how a wave propagates inside the electron gas. That will lead to a dispersion relation, which allows us to obtain the index of refraction. From the equations $q_\mu \tilde H_{\mu\nu}=0$ and $\epsilon_{\mu\nu\alpha\beta} q_\nu \tilde F_{\alpha\beta}=0$, combined with the constituent equations, one may proceed in the usual way to derive

\bea
&& [(\mu^{-1} + \frac{\omega}{|\vec{q}|} \sigma) |\vec{q}|^2 - (\epsilon - \frac{|\vec{q}|}{\omega} \tau) \omega^2] \tilde E_i=0, \\
&& [(\mu^{-1} +  \frac{\omega}{|\vec{q}|} \sigma) |\vec{q}|^2 - (\epsilon -  \frac{|\vec{q}|}{\omega} \tau) \omega^2] \tilde B_i=0.
\eea
For a plane-wave solution, the dispersion relation is

\be
\label{dispersion}
|\vec{q}|^2 - (\mu \epsilon)\, \omega^2 + 2(\mu \tau)\, \omega |\vec{q}| = 0.
\ee
The relation $|\vec{q}| = |\vec{q}|(\omega)$, which satisfies Eq. \eqref{dispersion}, leads to the index of refraction $n (\omega)= |\vec{q}|/\omega$. For $\tau=0$, we recover the usual expression $n = \sqrt{\mu}  \sqrt {\epsilon}$. Notice that, in the long-wavelength limit, one has $\tau=0$, $n = \sqrt{\mu}  \sqrt {\epsilon}$, and electric and magnetic responses that may be simultaneously negative. It then follows that one may obtain {\it negative indices of refraction} for the relativistic regime in such a limit.

In conclusion, we have calculated electromagnetic responses from quantum field theory at finite temperature and density for a relativistic electron gas. We have shown that, in the relativistic regime, the gas will exhibit Drude-type responses for both $\epsilon$ and $\mu^{-1}$ in the $|\vec{q}|\rightarrow 0$ long-wavelength limit, implying that such quantities may be simultaneously negative. The generalized formula [cf. Eq. \eqref{dispersion}] obtained for the index of  refraction for the electron plasma reduces to the $n = \sqrt{\mu}  \sqrt {\epsilon}$ usual one for $|\vec{q}|\rightarrow 0$, so that the plasma will exhibit a negative index of refraction under those conditions.

Summing up, the electrodynamics of materials with negative indices of refraction, with electrical permittivity and magnetic permeability {\it simultaneously} negative, was investigated by Veselago \cite{Veselago}, and remained a mathematical curiosity for many years, until split ring resonators enhanced magnetic responses to allow for negative permeabilities simultaneously with negative permittivities \cite{Smith}. Relativistic systems, however, do not suffer from a $v/c$ damping of magnetic effects with respect to electric ones. In fact, the last terms in Eqs. \eqref{responses1} and \eqref{responses2} dominate in the relativistic limit (which explains the $\omega_{m}=\sqrt{2} \,  \omega_{e}$ relation). Those are the terms that lead to Drude-type responses. As they are determined by symmetry, one may suggest that it is a {\it natural} behavior, which will occur for other relativistic systems, such as a gas of charged bosons. This problem is currently under investigation, and will be the object of a forthcoming article. Finally, this work suggests that the relativistic electron gas is one of nature's candidate for the realization of a negative index of refraction system. Our suggestion relies on the expectation that the RPA approximation, which does not take into account electron-electron interactions, is still a reasonable description of the electron gas (Appendix \ref{App1}), just as it is for simple metals. We are currently exploring our results numerically using physical parameters extracted from experimental data \cite{RGOC}. 

We do hope that present results will be of relevance for future studies in plasma physics, astrophysical observations, synchrotrons, and other environments with fast moving electrons. A specific experimental scenario where relativistic $T=0$ results can be tested is in a synchrotron accelerator. The electrons inside the beam are relativistic, well separated (which favors an indepent particle approximation), and one can actually probe the electromagnetic responses of the beam to externally applied time-dependent fields by monitoring the fields inside the beam. In astrophysical scenarios, relativistic electron gases do occur \cite{LN}, and their electromagnetic responses may be probed by comparing incident and scattered radiation. Finally, temperature effects may be tested in relativistic electron gases ejected from stars.

\acknowledgments

The author wishes to thank L. E. Oliveira for many fruitful discussions, for a critical reading, and for his warm hospitality at the Institute of Physics of the Universidade Estadual de Campinas, where part of this work was performed. Thanks are also due to F. S. S. Rosa and E. Reyes-G\' omez for discussions, comments, and suggestions, to CNPq and FAPESP for partial financial support, to H. Westfahl, for information about synchrotron beams, and to L. A. N. da Costa for suggestions on applications in astronomy.

\appendix
 
\section {Semiclassical expansion}
\label{App1}

In expression \eqref{Z}, we write $A_\mu=A^{(c)}_\mu + \hbar a_\mu$ ($\hbar=1$), where $A_\mu=A^{(c)}_\mu$ is a classical solution of the sourceless equation of motion for $A_\mu$, which we identify with the external classical field incident on the electron gas \cite{russo}. The Lagrangean for the EM field becomes
\be
\label{A1}
{\cal L}_A=-\frac{1}{4} F^{(c)}_{\mu\nu} F^{(c)}_{\mu\nu} - -\frac{1}{4} f_{\mu\nu} f_{\mu\nu}.
\ee
We then integrate over $a_\mu$ before doing the fermion integration
\be
\label{int a}
\oint  [da_\mu]  \det (\delta {\cal G}/\delta \Lambda) \, \delta({\cal G}) e^ {-S_{a}[a_\mu, \psi, \bar{\psi}]} 
\ee
with $S_a$ given by the quadratic form
\be
\label{Sa}
S_a=\frac{1}{2} \iint dx dy \,  a^x_{\mu} [G^\gamma_{\mu\nu}]^{-1}_{xy} a^y_{\nu} + e \int dx\,  (\bar{\psi} \gamma_\mu \psi)_x a^x_{\mu},
\ee
where we have used the shorthand $\int dx \equiv \int_0^\beta dx_4\int d^3 x$, and  $G^\gamma_{\mu\nu}$ is the photon propagator in the chosen gauge. The quadratic integral may be performed. Taking minus its logarithm
\be
\label{4fermi}
S_{e}^{int}=-\frac{e^2}{2} \iint dx dy (\bar{\psi} \gamma_\mu \psi)_x G^\gamma_{\mu\nu}(x-y) (\bar{\psi} \gamma_\nu \psi)_y.
\ee
The integral over quantum fluctuations of the gauge field leads to electron-electron interactions mediated by the photon propagator. The remaining fermionic integral is given by
\be
\label{fint}
Z^{(sc)}_e [A^{(c)}] = \oint [i d\psi^{\dagger}][d\psi] e^{- S^{(sc)}_{e} [\psi^{\dagger}, \psi, A^{(c)}]},
\ee
where the fermionic semiclassical action is $S_ e^{(sc)} = S_e + S_e^{int}$. Expanding $\exp(S_e^{int})$, the fermion integral reads
\be
\label{4fexp}
Z^{(sc)} _e [A^{(c)}] \cong \oint [i d\psi^{\dagger}][d\psi] e^{- S_{e} [\psi^{\dagger}, \psi, A^{(c)}]} [1+ S_e^{int}],
\ee
where we have neglected a term $O(\alpha^4)$. 

\begin{figure}
\epsfig{file=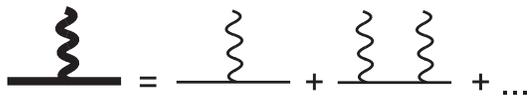,width=.8\columnwidth}
\caption{Expansion of the electron propagator in the external field, represented by wiggly lines.}
\label{fig1}
\end{figure}

The approximation in \eqref{3} only kept the leading term in \eqref{4fexp}. There, we dropped the superscript $c$ with the understanding that $A$ is a classical field.The fermion determinant which results from the integration involves the electron propagator in the presence of the background field. That propagator can be expanded in the background Fig. \ref{fig1}, so that $\hbox{Tr} \ln [-\beta \gamma_4 \Gamma_{A}] - \hbox{Tr} \ln [-\beta \gamma_4 G^{-1}_{0}]$ is given as an infinite sum of one-loop graphs: a fermion loop with an even number (due to Furry's theorem \cite{IZ}) of insertions of the classical field, 
\be
\label{one-loop}
 \frac{1}{2} \hbox{Tr} (G_0 \slashed{A} G_0 \slashed{A}) + \frac{1}{4} \hbox{Tr} (G_0 \slashed{A} G_0 \slashed{A} G_0 \slashed{A} G_0 \slashed{A})+ ....
\ee

The first term of the series is just
\be
\label{appol}
\frac{1}{\beta}\sum_n \int \frac{d^3q}{(2\pi)^3} {\tilde A}_\mu (q) {\tilde \Pi}_{\mu\nu}  (q) {\tilde A}_\nu (-q),
\ee
with ${\tilde \Pi}_{\mu\nu}  (q)$ given by \eqref{pol}, the one-loop vacuum polarization tensor \cite{IZ}. The next term, with four insertions, is still one-loop, nonlinear in the fields, depending on $(T,\xi)$, and typically of order $\alpha(\alpha E^2/m^4)$ or $\alpha(\alpha B^2/m^4)$.

\begin{figure}
\epsfig{file=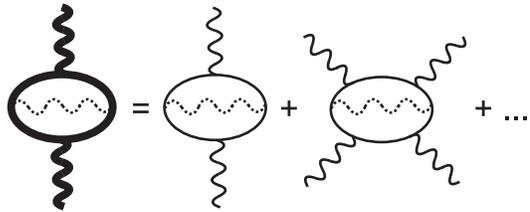,width=.8\columnwidth}
\caption{Graph for the e-e interaction expanded in the external field (wiggly lines) The dashed wiggly lines represent the photon propagator.}
\label{fig2}
\end{figure}

If we consider the first contribution from the e-e intercation, we have to contract the four fermion term in $S_e^{int}$ with the electron propagator in the external field. The resulting graph (Fig. \ref{fig2}) is a two-loop contribution. When we expand in the external field, the first contribution that depends on the field is quadratic and of order $\alpha^2$, and contributes in linear response. Next terms in the expansion in the external field are nonlinear $(T, \xi)$-dependent contributions of order $\alpha(\alpha E^2/m^4)$, $\alpha(\alpha B^2/m^4)$. 

In conclusion, restricting our attention to formula \eqref{3} is equivalent to neglecting one-loop contributions that are nonlinear, as well as a two-loop contribution to linear response of order $\alpha^2$, and nonlinear ones that also come in with e-e interactions. Although nonlinear terms might bring in interesting effects \cite{refB}, we will restrict our analysis to fields that are not strong enough to invalidate the linear response approximation. Therefore, we only consider the interaction of independent electrons with weak external fields. 
 
\section{Relativistic Drude expressions}
\label{App2}

Equations \eqref{calA} and \eqref{calB} may be integrated over angles and continued to Minkowski space $q_4 \rightarrow i\omega$ to give
\be
{\cal B}^\ast= \frac{-e^2}{\pi^2 q_M^2} \int_0^\infty \frac{dp \, p^2 n_F}{\omega_p} [1+ \frac{4\omega_p^2+q_M^2}{8p|\vec{q}|} f_1 - \frac{\omega_p\omega}{2p|\vec{q}|} f_2],
\ee
\be
{\cal D}^\ast = \frac{-e^2}{\pi^2 q_M^2} \int_0^\infty \frac{dp \, p^2 n_F}{\omega_p} [1+ \frac{2m^2+q_M^2}{8p|\vec{q}|} f_1],
\ee
where ${\cal D}^\ast \equiv {\cal A}^\ast - (1+ \frac{3q_M^2}{2|\vec{q}|^2}){\cal B}^\ast$, and  $q_M^2=\omega^2- |\vec{q}|^2$. The functions $f_1$ and $f_2$ are
\be
f_1= \ln \frac{(q_M^2-2p|\vec{q}|)^2 - 4 \omega_p^2\omega^2}{(q_M^2+2p|\vec{q}|)^2 - 4 \omega_p^2\omega^2},
\ee
\be
f_2= \ln \frac{\omega^4- 4 (\omega\omega_p+p|\vec{q}|)^2}{\omega^4- 4 (\omega\omega_p-p|\vec{q}|)^2}.
\ee
Introducing the dimensionless variables $x\equiv \omega_p/m$, $a\equiv \omega/2m$, and $b\equiv |\vec{q}|/2m$, and the functions
\be
\label{L1}
L_1(a,b)\equiv \ln(ax+b\sqrt{x^2-1}+a^2-b^2),
\ee
\be
\label{L2}
L_2(a,b) \equiv \ln(ax+b\sqrt{x^2-1}+a^2),
\ee
we may rewrite
\be
\label{log1}
f_1= - L_1(a,b)-L_1(-a,b)+L_1(a,-b)+L_1(-a,-b)
\ee
\be
\label{log2}
f_2= +L_2(a,b)+L_2(-a,-b)-L_2(a,-b)-L_2(-a,b),
\ee
These integrals may be performed for any values of $a$ and $b$ \cite{BCQ}. In \cite{RGOC}, we have numerically verified that the extrapolation of relativistic expressions for $\epsilon$ (after angular integration) to nonrelativistic  parameters fits well the experimental data for plasmon frequencies in condensed matter systems at finite temperature. This is because we have Drude expressions in both the relativistic and nonrelativistic regimes. That is not the case for $\mu^{-1}$, where a Drude expression only appears in the relativistic regime. 

However, as we are interested in the long wavelength limit, $|\vec{q}|\rightarrow 0$, we expand $f_1$ and $f_2$ in powers of $b$
\be
\label{expansion1}
\frac{f_1}{\sqrt{x^2-1}}=-\frac{2b}{a}  F_-^{(1)}-\frac{2b^3}{3a^3}[F_-^{(1)}+aF_+^{(2)}+(a^2-1)F_-^{(3)}],
\ee
\be
\label{expansion2}
\frac{f_2}{\sqrt{x^2-1}}=+\frac{2b}{a}  F_+^{(1)}+\frac{2b^3}{3a^3}[F_+^{(1)} - 2aF_-^{(2)}+(a^2-1)F_+^{(3)}],
\ee
where we have used
\be
\label{Fj}
F_{\pm}^{(j)}\equiv \frac{1}{(x+a)^j} \pm \frac{1}{(x-a)^j}.
\ee
In terms of the dimensionless variables introduced above, we have
\bea
{\cal B}^\ast&=&-\frac{e^2}{4\pi^2}\frac{1}{a^2- b^2}\int_1^\infty dx\, n_F(x) [\sqrt{x^2-1}\nonumber \\ 
&+& \frac{(x^2+a^2-b^2)}{4b} f_1 - \frac{2a}{4b}f_2]
\eea
\bea
{\cal D}^\ast&=&-\frac{e^2}{4\pi^2}\frac{1}{a^2- b^2}\int_1^\infty dx\, n_F(x) [\sqrt{x^2-1}\nonumber \\ 
&+& \frac{(1+2a^2-2b^2)}{8b} f_1]
\eea
Using \eqref{expansion1} and \eqref{expansion2}, we obtain
\be
\label{Basterisk}
\frac{a^2}{b^2}{\cal B}^\ast=\frac{e^2}{4\pi^2}\left[\frac{2}{3a^2} I^{(0)}+\frac{1+14a^2}{3a^2} I^{(1)}+ 4a^2 I^{(2)}\right],
\ee
\be
a^2{\cal D}^\ast=-\frac{e^2}{4\pi^2}\left[I^{(0)}+\frac{1+2a^2}{2} I^{(1)}\right],
\ee
which lead to
\be
\label{Aasterisk}
{\cal A}^\ast=-\frac{3e^2}{2\pi^2}\left[I^{(1)}+ a^2 I^{(2)}\right],
\ee
where the integrals $I^{(j)}$, related to \eqref{Fj}, are given by
\be
\label{Ij}
I^{(j)}(a^2)\equiv \int_1^\infty dx\, n_F(x) \frac{\sqrt{x^2-1}}{(x^2-a^2)^j},
\ee
with $I^{(2)}=\partial I^{(1)}/\partial a^2$.

We may compute these integrals exactly at $T=0$, when $n_F(x)=\Theta(\zeta -x)$. We use the Euler substitution $\sqrt{(x-1)(x+1)}=t(x+1)$ and decomposition in partial fractions to derive
\bea
\label{integralsI}
&& I^{(0)}=\frac{1}{2}[\zeta\sqrt{\zeta^2-1} - \ln(\zeta+\sqrt{\zeta^2-1})], \\
&& I^{(1)}= \ln(\zeta+\sqrt{\zeta^2-1)}- \frac{1}{\sigma(a)}\mathrm{arctg}(\frac{\sigma(a)}{\sigma(\zeta)},
\eea
where $\sigma (y) \equiv y/\sqrt{|1-y^2|}$. We have used $q_M^2\rightarrow \omega^2 > 0$. Since we will be interested in $\omega \rightarrow 0$, we have also taken $a<<1$. 

Using the expressions
\bea
\label{epsfinal}
&& \epsilon= 1+  {\cal C}^\ast+ {\cal A}^\ast + (1-\frac{a^2}{b^2}) {\cal B}^\ast , \\
&&  \mu^{-1}= 1+ 2 {\cal C}^\ast+ {\cal A}^\ast - 2 \frac{a^2}{b^2} {\cal B}^\ast,  
\label{mufinal}
\eea
and expanding \eqref{Basterisk} and \eqref{Aasterisk} for $a<<1$ (${\cal C}^\ast$ is $O(a^2)$), we obtain
\bea
\label{Drudeeps}
&& \epsilon=1 - \frac{a_e^2}{a^2} + \frac{e^2}{3\pi^2} g_e (\zeta) + O(a^2), \\
&& \mu^{-1}=1 - \frac{a_m^2}{a^2} - \frac{5e^2}{6\pi^2} g_m (\zeta) + O(a^2),
\label{Drudemu}
\eea
where $a^2_m=2a^2_e$,
\be
\label{Drudefreq}
a^2_e=\frac{\omega^2_e}{4m^2}= \frac{e^2}{12\pi^2} \frac{(\zeta^2-1)^{3/2}}{\zeta},
\ee
and the $O(\alpha)$ corrections are given by
\bea
&& g_e(\zeta)= \ln(\zeta+\sqrt{\zeta^2-1})-\frac{1}{\sigma(\zeta)}-\frac{7}{6\sigma^{3}(\zeta)}, \\
&& g_m(\zeta)= \ln(\zeta+\sqrt{\zeta^2-1})-\frac{1}{\sigma(\zeta)}-\frac{14}{15\sigma^{3}(\zeta)}.
\eea


\begin{references}
\bibitem{MtoE} Euclidean and Minkowski coordinates are related by $x_4= i x^0$ and $x_j=x^j$. For gauge fields, $A_4=iA^0$ and $A_j=A^j$. For Dirac matrices, $\gamma_4=\gamma^0$ and $\gamma_j=\gamma^j$. The Minkowski metric is $\eta_{\mu\nu}=(+,-)$.
\end{references}
\end{document}